\documentclass[a4paper,11pt]{article}
\usepackage{pos}

\newcommand{\dx}[1]{\hspace{-0.4em}\ensuremath{\mathrm{d}#1}\,}

\newcommand{\eqn}[1]{Eq.~(\ref{#1})}
\newcommand{\fig}[1]{Fig.~\ref{#1}}
\newcommand{\tab}[1]{Table~\ref{#1}}

\title{ 
 \vspace{-25mm}
  \begin{flushright}
    LFTC-21-2/63
  \end{flushright}
  \vspace{15mm}
  Charmonium in nuclear matter and nuclei}
  \ShortTitle{Charmonium in nuclear matter and nuclei}

\author*[a]{J.J.~Cobos Mart\'inez}
\author[b]{K.~Tsushima}
\author[c]{G.~Krein}
\author[d]{A.W.~Thomas}

\affiliation[a]{Departamento de F\'isica, Universidad de Sonora, Boulevard
Luis Encinas J. y Rosales, Colonia Centro, Hermosillo, Sonora 83000, M\'exico}
\affiliation[b]{Laborat\'orio de F\'{\i}sica Te\'orica e Computacional-LFTC, 
Universidade Cidade de S\~ao Paulo,
01506-000, S\~ao Paulo, SP, Brazil}
\affiliation[c]{Instituto de F\'{\i}sica Te\'orica, Universidade Estadual 
Paulista, Rua Dr. Bento Teobaldo Ferraz, 271 - Bloco II, 01140-070, 
S\~ao Paulo, SP, Brazil}
\affiliation[d]{CSSM, School of Physical Sciences, University of Adelaide,
Adelaide SA 5005, Australia}

\emailAdd{jesus.cobos@fisica.uson.mx}
\emailAdd{kazuo.tsushima@gmail.com}
\emailAdd{gkrein@ift.unesp.br}
\emailAdd{anthony.thomas@adelaide.edu.au}

\abstract{We present results for the $\eta_c$-nucleus bound state energies for various nuclei using 
an effective Lagrangians approach. The attractive potentials for the $\eta_c$ in the nuclear medium
originate from the medium-modified intermediate $D D^{*}$ state in the $\eta_c$ self energy,
using the local density approximation. Our results suggest that the $\eta_c$  should form bound states with all the nuclei considered.}

\FullConference{%
  *** 10th International Workshop on Charm Physics (CHARM2020), ***\\
  *** 31 May - 4 June, 2021 ***\\
  *** Mexico City, Mexico - Online ***
}

\begin{document}
\maketitle

\section{Introduction}

Quantum chromodynamics (QCD) is the accepted fundamental theory  of the strong interactions. However, a quantitative understanding of the strong force and strongly
interacting matter in vacuum and extreme conditions of temperature and density from the  underlying
theory is still limited. Even though there are numerous reasons to study charmonium states, the study 
of the interactions of these states with atomic nuclei offers potential  opportunities to gain an
 understanding of the workings  of strongly interacting matter and QCD. Since charmonium and 
 nucleons do not share light ($u,\,d$) quarks, the OZI rule suppresses the interactions mediated by 
 the exchange of  mesons made of only light quarks.  If such states are indeed bound to nuclei,  it is
  therefore important to search for other sources of attraction which could lead to the binding of charmonia to atomic nuclei.  
Since the suggestion of Brodsky~\cite{Brodsky:1989jd}, more than three decades ago, that charmonium
 states may be bound with nuclei, a large amount of research looking for alternatives to the meson
  exchange mechanism has accumulated over the years to investigate the possible existence of such 
 exotic states~\cite{Krein:2017usp}. 
 The discovery of such bound states would represent an important step forward in our 
 understanding of the nature of strongly interacting systems.

There is a great amount of evidence that the internal structure of hadrons changes in medium and this
must be taken into account when addressing  charmonium in nuclei. 
Studies carried by some of us~\cite{Krein:2010vp,Tsushima:2011kh} have shown that the effect of the
 nuclear mean fields on the intermedium $D\overline{D}$ state is of particular importance in the $J/\Psi$
  case; the modifications induced by the strong nuclear mean  fields on the $D$ mesons' light-quark
 content enhance the self-energy in  such a way  as to provide an attractive $J/\Psi-$nucleus effective
  potential. 
We extended this approach to the case of $\eta_c$  in Ref.~\cite{Cobos-Martinez:2020ynh} by
 considering an intermediate $DD^{*}$ state in the $\eta_c$ self-energy with the ligth quarks created 
 from the vacuum. Here the effective scalar and vector meson 
mean fields  in the nuclear medium couple to the light $u$  and $d$ quarks in the charmed 
mesons and this, as in the case of the $J/\Psi$, provides attraction to the $\eta_c$ in the nuclear
medium. Recently, we have extended this approach to the case of bottomonium~\cite{Zeminiani:2020aho}.

This paper is organized as follows. In Sec.~\ref{sec:mass_shifts} we briefly discuss the computation and
 present  results for the mass shift  of the $\eta_c$ in symmetric nuclear matter.
Using the results of Sec.~\ref{sec:mass_shifts}, together with the density profiles of the nuclei studied,
 in Sec.~\ref{sec:etac_bound_states}  we present results for the scalar $\eta_c$-nucleus potentials,  as well as the corresponding bound state energies. Finally, in Sec.~\ref{sec:summary} we present our
 conclusions.

\section{\label{sec:mass_shifts}$\eta_c$ in nuclear matter}

\begin{figure}
\begin{minipage}{0.45\textwidth}
\includegraphics[scale=2.20]{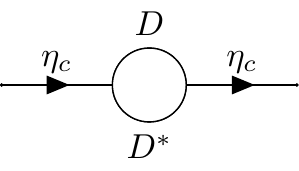}
\end{minipage}
\begin{minipage}{0.55\textwidth}
\includegraphics[scale=0.275]{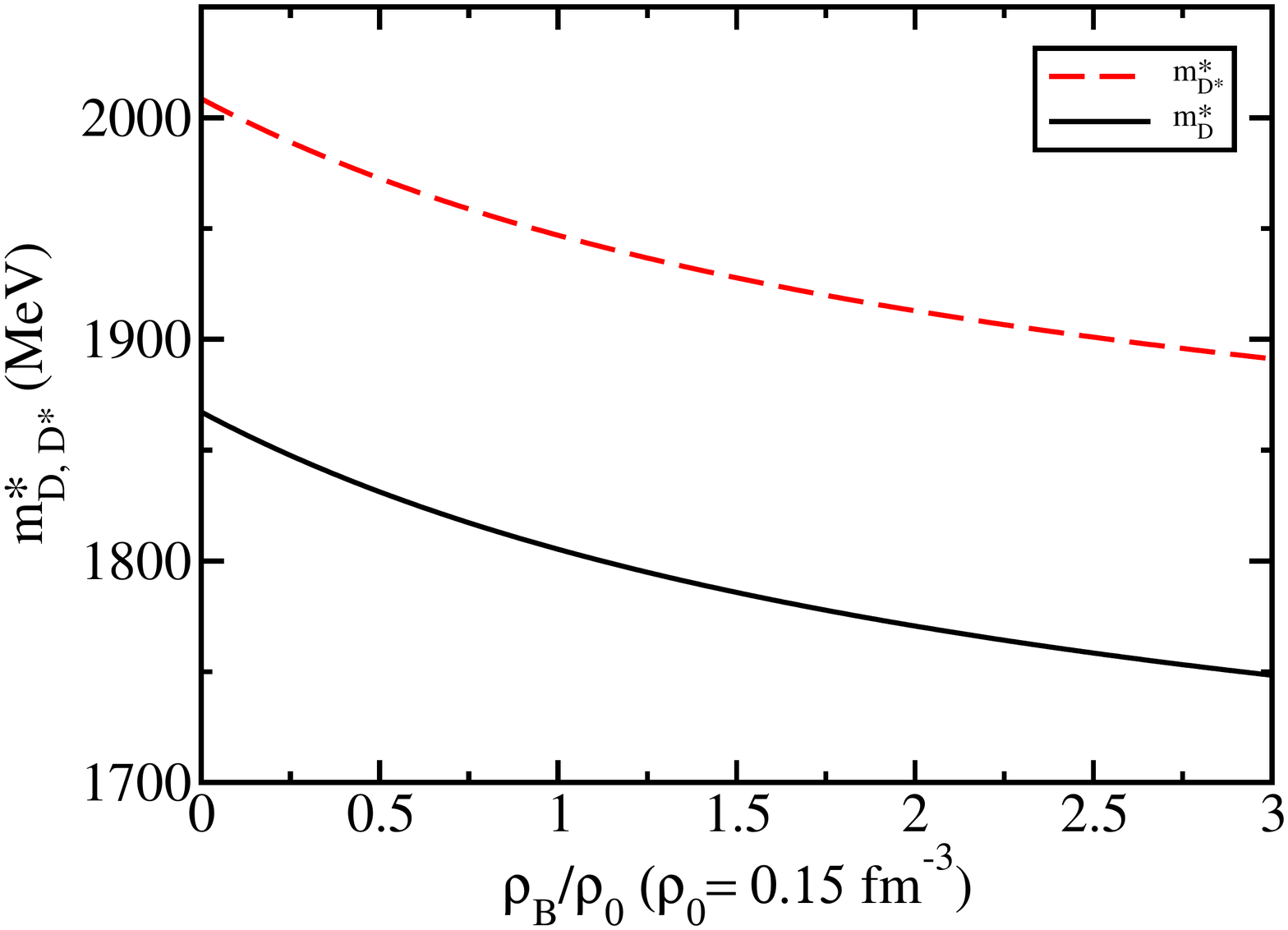}
\end{minipage}
\caption{\label{fig:DDs} Left panel: Intermediate $DD^{*}$ state contribution to the 
$\eta_c$ self energy. Right panel: In-medium $D$ and $D^{*}$ meson masses in nuclear matter.}
\end{figure}

In computing the $\eta_c$ mass shift in nuclear matter we take into account only the 
intermediate $DD^{*}$ state contribution to the $\eta_c$ self-energy; see Refs~\cite{Cobos-Martinez:2020ynh,Zeminiani:2020aho} and references therein  for details. 
The effective interaction Lagrangian for the $\eta_c D D^{*}$ vertex is given by 
\begin{eqnarray}
\label{eqn:LetacDDast}
{\cal L}_{\eta_c DD^{*}} = i g_{\eta_c DD^{*}}
\left[(\partial^\mu \eta_c) \left( \overline{D}^{*}_\mu D - \overline{D} D^{*}_\mu \right)
- \eta_c\left( \overline{D}^{*}_\mu (\partial^\mu D) - (\partial^\mu \overline{D}) D^{*}_\mu \right)
\right], 
\end{eqnarray}
where $D^{(*)}$ represents the $D^{(*)}$-meson field isospin doublet, and $g_{\eta_c D D^{*}}$
 is the coupling constant. Using \eqn{eqn:LetacDDast}, the Feynman diagram  of \fig{fig:DDs}, 
 and considering an $\eta_c$ at rest, the $\eta_c$ self-energy is given by, 
\begin{equation}
    \label{eqn:etac_se}
    \Sigma_{\eta_c}(k^{2})= \frac{8g_{\eta_c D D^{*}}^{2}}{\pi^{2}}\int_{0}^{\infty}
    \dx{k}k^{2}I(k^{2})
\end{equation}
where $I(k^2)$ is given by
\begin{align}
I(k^{2})&= \left. \frac{m_{\eta_c}^{2}(-1+k^{0\,2}/m_{D^{*}}^{2})}
{(k^{0}+\omega_{D^{*}})(k^{0}-\omega_{D^{*}}) 
(k^{0}-m_{\eta_c}-\omega_{D})}\right|_{k^{0}=m_{\eta_c}-\omega_{D^{*}}} 
\nonumber \\
+&\left. \frac{m_{\eta_c}^{2}(-1+k^{0\,2}/m_{D^{*}}^{2})}
{(k^{0}-\omega_{D^{*}})(k^{0}-m_{\eta_c}+\omega_{D}) 
(k^{0}-m_{\eta_c}-\omega_{D})}\right|_{k^{0}=-\omega_{D^{*}}},  
\end{align}
and $\omega_{D^{(*)}}=(k^{2}+m_{D^{(*)}}^{2})^{1/2}$, with $k=|\vec{k}|$.
The integral in \eqn{eqn:etac_se} is divergent and will be regulated with a phenomenological
vertex form factor, as in Refs~\cite{Krein:2010vp, Tsushima:2011kh, 
Cobos-Martinez:2020ynh, Zeminiani:2020aho,Cobos-Martinez:2017vtr,Cobos-Martinez:2017woo},
\begin{equation}
    \label{eqn:FF}
    u_{D^{(*)}}(k^{2})=
    \left(\frac{\Lambda_{D}^{2} + m_{\eta_c}^{2}}
    {\Lambda_{D}^{2}+4\omega_{D^{(*)}}^{2}(k^{2})}
    \right)^{2},
\end{equation}
with cutoff parameter $\Lambda_{D}\,(=\Lambda_{D^{*}})$, by introducing the  factor $u_{D}(k^{2})u_{D^{*}}(k^{2})$
 into the integrand of \eqn{eqn:etac_se} .
The cutoff parameter $\Lambda_D$ is an unknown input to our calculation. Its value has been determined
phenomenologically in Ref~\cite{Krein:2010vp} to be $\Lambda_{D}\approx 2500\,\textrm{MeV}$. 
However, in order to quantify the uncertainity of our results to its value we present results for 
$\Lambda_D$ in the range 1500-3000 MeV.
%
%
\begin{figure*}
\centering
\includegraphics[scale=0.275]{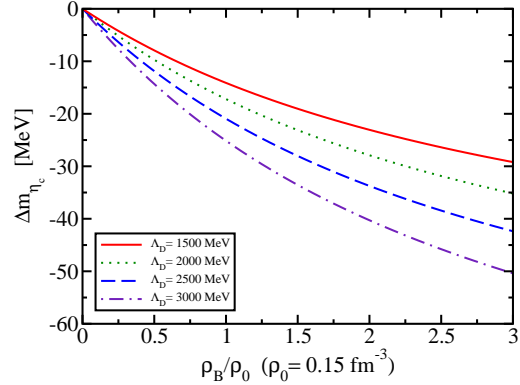}
\caption{\label{fig:etac_mass_shift} $\eta_c$ mass shift as a function of the nuclear matter density for various
 values of the cutoff parameter.}
\end{figure*}

%
%
The $\eta_c$ mass shift in nuclear matter is computed from the difference between the in-medium,
$m_{\eta_c}^{*}$, and vacuum, $m_{\eta_c}$, masses of the $\eta_c$,
\begin{equation}
    \Delta m_{\eta_c}= m_{\eta_c}^{*}-m_{\eta_c},
\end{equation}
with the masses computed by solving
\begin{equation}
   \label{eqn:etac_mass}
    m_{\eta_c}^{2}= (m_{\eta_c}^{0})^{2} + \Sigma_{\eta_c}(k^{2}=m_{\eta_c}^{2}),
\end{equation}
where $m_{\eta_c}^{0}$ is the bare $\eta_c$ mass and $\Sigma_{\eta_c}(k^{2})$ is given by 
\eqn{eqn:etac_se}.  The $\eta_c$ meson bare mass  is fixed using \eqn{eqn:etac_mass} by fitting the
physical $\eta_c$ mass, $m_{\eta_c}=2983.9$ MeV. 
(For the other parameters in vacuum we use 
$g_{\eta_c D D^{*}}=(0.6/\sqrt{2})g_{J/\Psi DD}=3.24$~\cite{Lin:1999ad,Lucha:2015dda}, 
$m_{D}=1867.2$ MeV, and  $m_{D^{*}}=2008.6$ MeV; see Ref.~\cite{Cobos-Martinez:2020ynh} 
for details). 
The $\eta_c$ mass in the nuclear medium is similarly computed from \eqn{eqn:etac_mass} with the 
self-energy calculated with the medium-modified $D$ and $D^{*}$ masses.
The nuclear density dependence of the $\eta_c$ mass originates from the interactions of the
intermediate  $DD^{*}$ state with the nuclear medium through their medium-modified masses,
$m_{D}^{*}$ and $m_{D^{*}}^{*}$, respectively. These in-medium masses  are  calculated within
 the quark-meson  coupling (QMC) model~\cite{Krein:2010vp,Tsushima:2011kh}, in which effective 
 scalar and vector  meson mean fields couple to the light quarks in the charmed 
 mesons~\cite{Krein:2010vp, Tsushima:2011kh}.
The resulting medium-modified masses for the $D$ and $D^{*}$ mesons are shown in the left
panel of  \fig{fig:DDs} as a function of $\rho_{B}/\rho_{0}$, where $\rho_{B}$ is the baryon 
density of nuclear matter and  $\rho_{0}=0.15$ fm$^{-3}$ is the  saturation density of symmetric 
nuclear  matter. As can be seen from  \fig{fig:DDs} the masses of the $D$ and $D^{*}$ mesons
are reduced in the nuclear medium. 
In \fig{fig:etac_mass_shift}, we present the $\eta_c$ mass shift, $\Delta m_{\eta_c}$, as a function of  
$\rho_{B}/\rho_{0}$ for several values of $\Lambda_{D}$. 
As can be seen from \fig{fig:etac_mass_shift}, the effect of the nuclear medium is to shift the 
$\eta_c$ mass  downwards.  This happens for all values of $\Lambda_D$.

\section{\label{sec:etac_bound_states}$\eta_c$ in nuclei}

\begin{figure*}
\begin{tabular}{cc}
\includegraphics[scale=0.275]{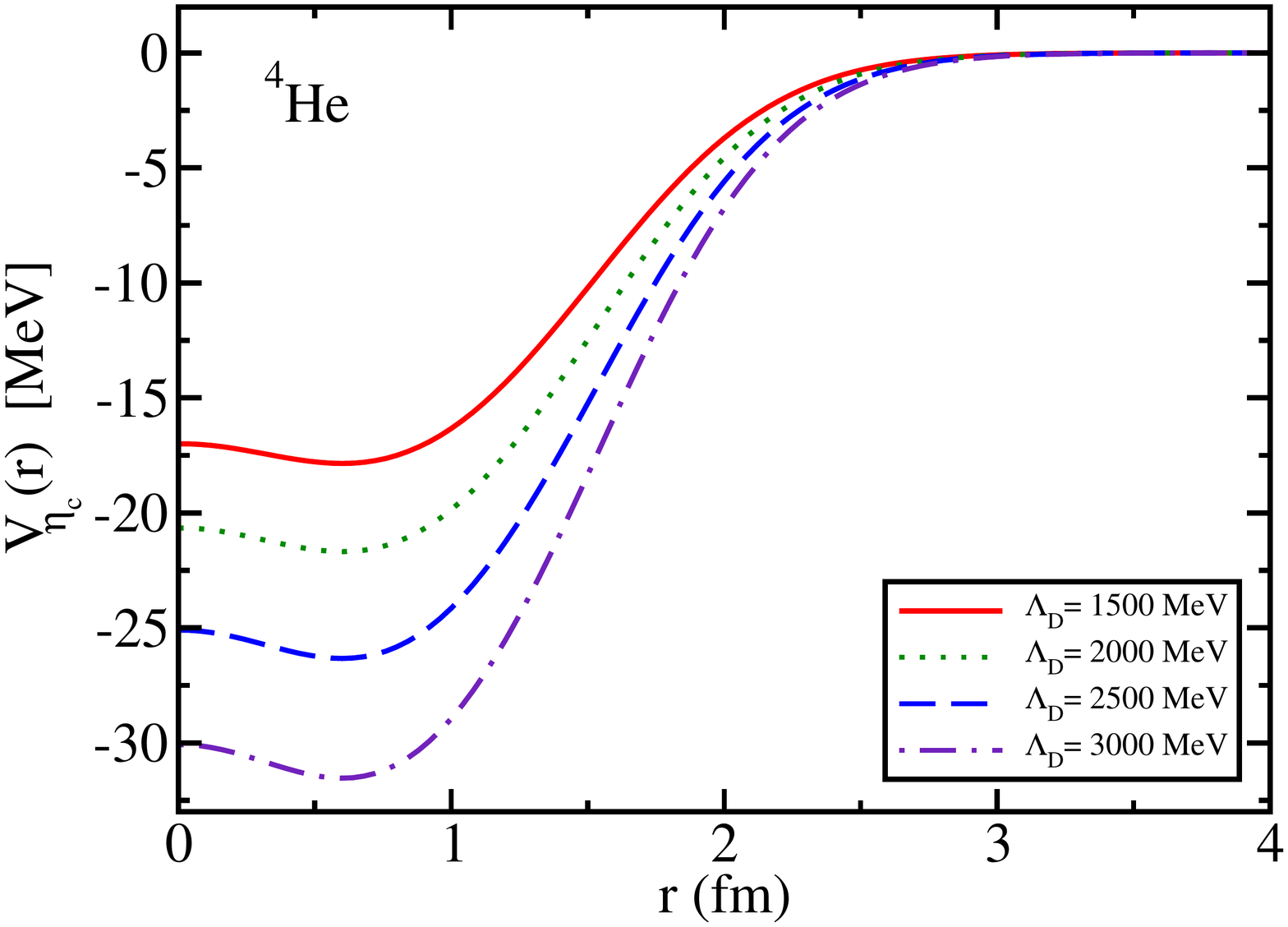} &  
\includegraphics[scale=0.275]{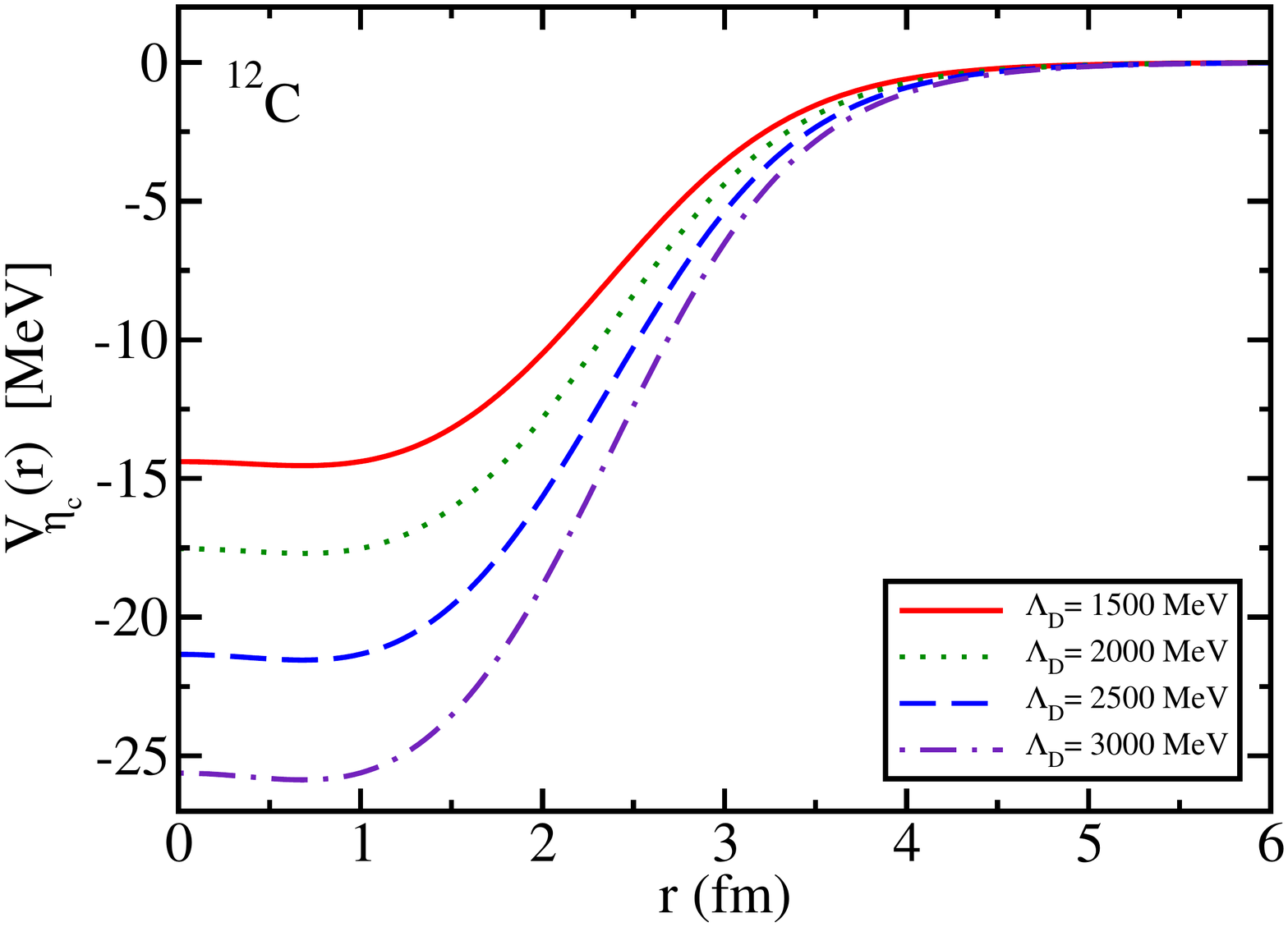}  \\
\includegraphics[scale=0.275]{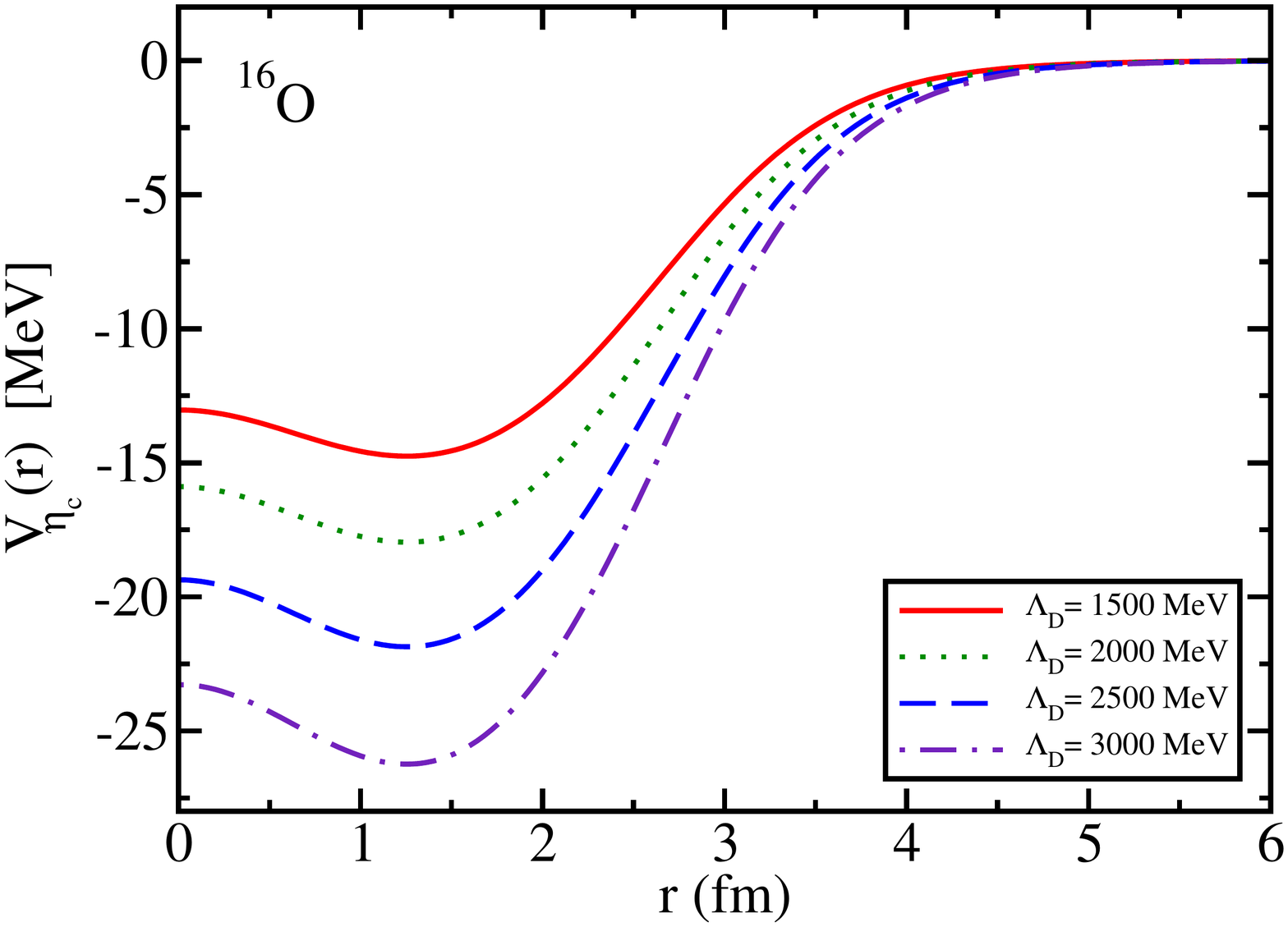} &
\includegraphics[scale=0.275]{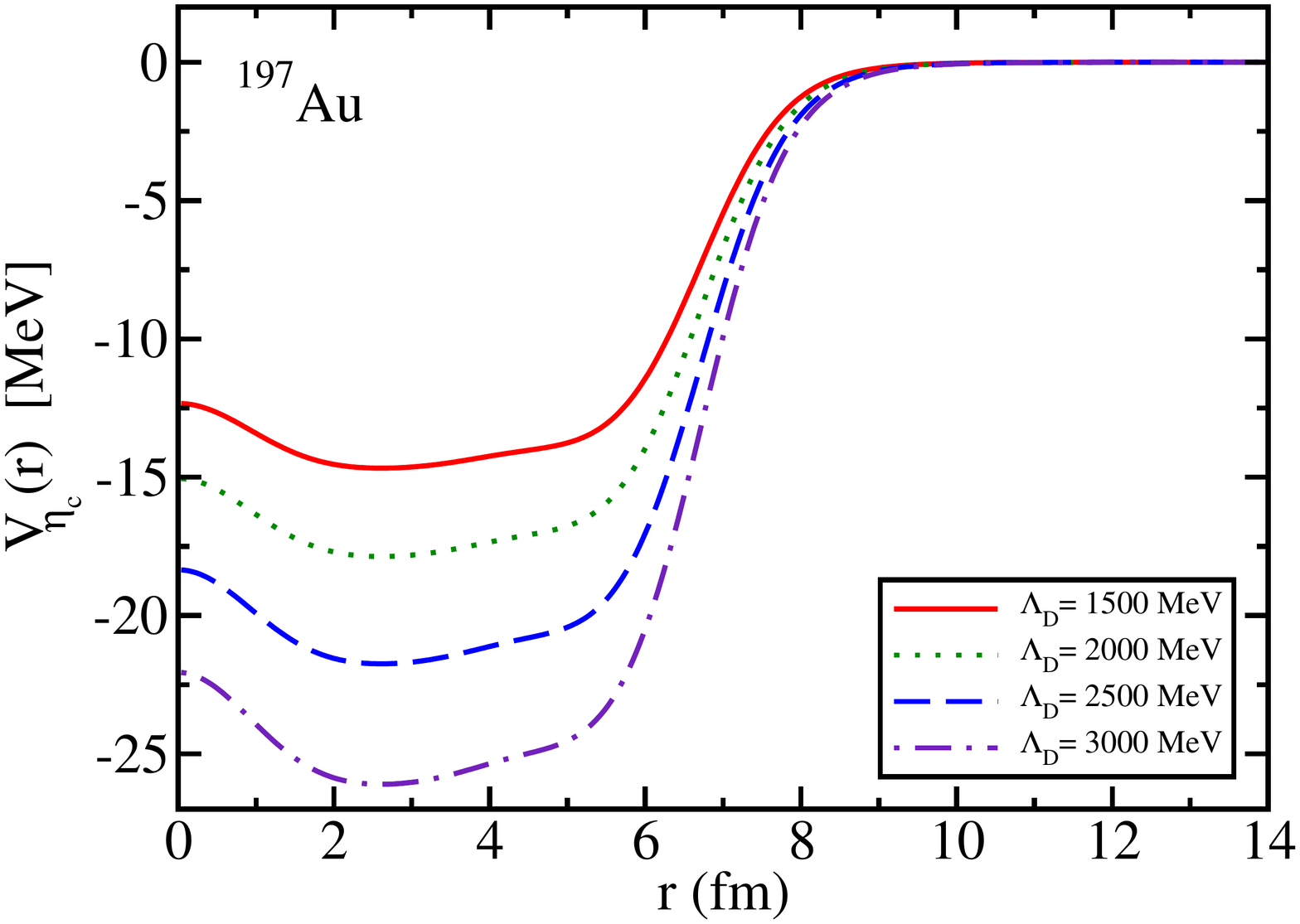}  \\
\includegraphics[scale=0.275]{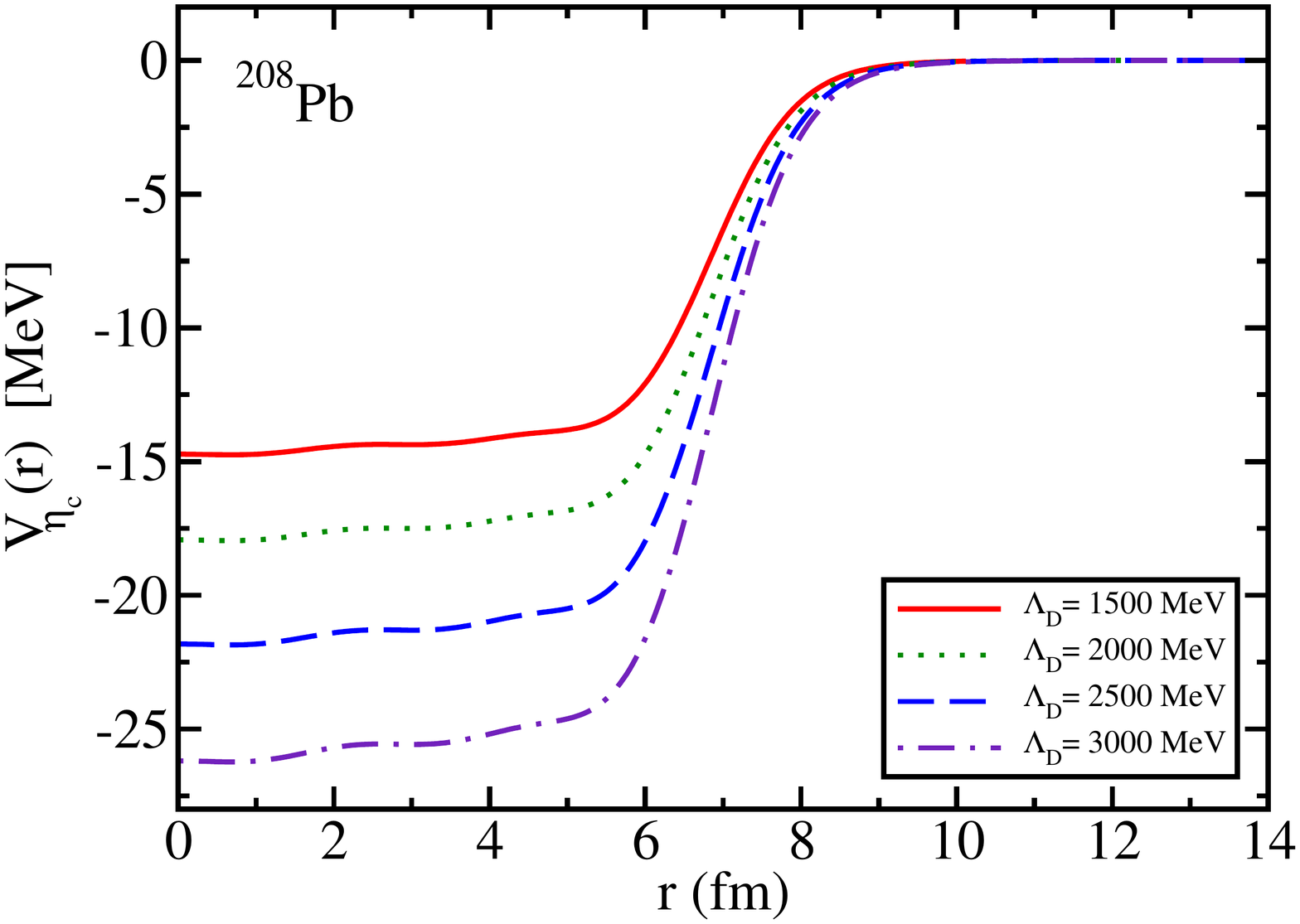}   & \\
  \end{tabular}
  \caption{\label{fig:etac_potentials}
    $\eta_c$-nucleus  potentials for various nuclei and values of the cutoff parameter $\Lambda_{D}$.}
\end{figure*}

A negative mass shift for the $\eta_c$ means that the nuclear medium provides attraction to the 
$\eta_c$ and opens the possibility for the binding of $\eta_c$ mesons to nuclei. Thus, we now 
consider the binding of the $\eta_c$ to nuclei when the $\eta_c$ is produced  inside a nucleus 
$A$ with baryon density distribution $\rho_{B}^{A}(r)$; see  Ref.~\cite{Cobos-Martinez:2020ynh} for 
a more detailed discussion.
Here we only consider the nuclei $^{4}$He, $^{12}$C, $^{16}$O, $^{197}$Au, and $^{208}$Pb,
whose baryon density distributions were also calculated within the  QMC  
model~\cite{Cobos-Martinez:2020ynh}, except that for $^{4}$He~\cite{Saito:1997ae}. Using a local density approximation, the $\eta_c$-meson potential  inside nucleus $A$ is given by 
\begin{equation}
\label{eqn:etac_potentials}
V_{\eta_c A}(r)= \Delta m_{\eta_c}(\rho_{B}^{A}(r)),
\end{equation}
\noindent where $ \Delta m_{\eta_c}$ mass shift shown in \fig{fig:etac_mass_shift}, and  $r$ is the distance 
from the center of the nucleus.
In~\fig{fig:etac_potentials} we present the $\eta_c$ potentials for the nuclei mentioned above and 
various values of $\Lambda_{D}$. From \fig{fig:etac_potentials} one can see that all $\eta_c$ potentials
 are attractive but their depths depend on $\Lambda_{D}$, being deeper for larger $\Lambda_{D}$. 

%
\begin{table}[ht!]
\begin{center}
\scalebox{0.75}{
\begin{tabular}{ll|r|r|r|r}
  \hline \hline
  & & \multicolumn{4}{c}{Bound state energies} \\
  \hline
& $n\ell$ & $\Lambda_{D}=1500$ & $\Lambda_{D}=2000$ & $\Lambda_{D}=2500$ & 
$\Lambda_{D}= 3000$ \\
\hline
$^{4}_{\eta_{c}}\text{He}$
    & 1s & -1.49 & -3.11 & -5.49 & -8.55 \\
\hline
$^{12}_{\eta_{c}}\text{C}$
    & 1s & -5.91 & -8.27 & -11.28 & -14.79 \\
    & 1p & -0.28 & -1.63 & -3.69  & -6.33 \\
\hline
$^{16}_{\eta_{c}}\text{O}$
    & 1s & -7.35 & -9.92 & -13.15 & -16.87 \\
    & 1p & -1.94 & -3.87 & -6.48  & -9.63 \\
\hline
$^{197}_{\eta_{c}}\text{Au}$
      & 1s & -12.57 & -15.59 & -19.26 & -23.41 \\
      & 1p & -11.17 & -14.14 & -17.77 & -21.87 \\
      & 1d & -9.42  & -12.31 & -15.87 & -19.90 \\
      & 2s & -8.69  & -11.53 & -15.04 & -19.02 \\
      & 1f & -7.39  & -10.19 & -13.70 & -17.61 \\
\hline
$^{208}_{\eta_{c}}\text{Pb}$
    & 1s & -12.99 & -16.09 & -19.82 & -24.12 \\
    & 1p & -11.60 & -14.64 & -18.37 & -22.59 \\
    & 1d & -9.86  & -12.83 & -16.49 & -20.63 \\
    & 2s & -9.16  & -12.09 & -15.70 & -19.80 \\
    & 1f & -7.85  & -10.74 & -14.30 & -18.37 \\
\hline 
\hline
\end{tabular}
}
  \caption{\label{tab:etac-A-kg-be} $\eta_c$-nucleus bound state energies for various nuclei. 
  All dimensionful quantities are in MeV.}
\end{center}
\end{table}

Now we calculate the $\eta_c$--nucleus bound state energies for the nuclei listed above by solving the Klein-Gordon equation
\begin{equation}
\label{eqn:kg}
\left(-\nabla^{2} + m^{2} + 2 m V(\vec{r})\right)\phi_{\eta_c}(\vec{r})
= \mathcal{E}^{2}\phi_{\eta_c}(\vec{r}),
\end{equation}
\noindent where $m$ is the reduced mass of the $\eta_c$--nucleus system  in vacuum  and
$V(\vec{r})$ is the $\eta_c$-nucleus potential given in  \eqn{eqn:etac_potentials}. 

The computed bound state energies ($E$) of the $\eta_c$-nucleus system, given by 
$E= \mathcal{E}-m$, where  $\mathcal{E}$ is the energy eigenvalue in \eqn{eqn:kg}, are listed in
 \tab{tab:etac-A-kg-be} for four values of $\Lambda_{D}$.
The results in  \tab{tab:etac-A-kg-be} show that the $\eta_c$-meson is expected to form bound states with all the nuclei considered, independently of the value of $\Lambda_D$.
Clearly, however, particular values for the bound state energies are dependent on 
$\Lambda_{D}$. This dependence was expected since the $\eta_c$ potentials are also dependent on
$\Lambda_{D}$, and it is therefore an uncertainty in the results obtained in our approach.
Finally, note that  the binding is stronger for larger values of $\Lambda_{D}$ and that the $\eta_c$ 
is predicted to be bound more strongly to heavier nuclei.

\section{\label{sec:summary}Summary and Conclusions}

We have calculated the $\eta_c$--nucleus bound states energies for various nuclei.  
The $\eta_c$--nucleus potentials were calculated in the a local density approximation from the
$\eta_c$ mass shift as a function of the nuclear matter density and the baryon density profiles of the
nuclei studied. Using these potentials, we have solved the Klein-Gordon equation to obtain 
$\eta_c$--nucleus bound state energies. Our results show that one should expect the $\eta_c$
to form bound states for all the nuclei studied, even though  the precise values of the bound state
energies are dependent on the cutoff mass values $\Lambda_D$ used in the form factors. 
The discovery of such bound states  would represent an important step forward in our understanding 
of the  nature of strongly interacting systems. 
Note that we have ignored the natural width of 32 MeV in free space of the  $\eta_c$ and this could be 
an issue related to the observability of the predicted bound states. However, it should still be possible to 
see that there are bound states, which is the main result of this work. 
Addition of an imaginary part to the $\eta_c$ potentials is underway and will be reported 
elsewhere~\cite{bottomonium-nuclear-bs}
 Furthermore, we have done an inital study~\cite{Zeminiani:2020aho} using a different form factor since this may impact the results reported here.

\section{Acknowledgments}

This work was partially supported by Conselho Nacional de Desenvolvimento 
Cient\'{i}fico e Tecnol\'{o}gico (CNPq), process Nos.~313063/2018-4 (KT), 
426150/2018-0 (KT) and 309262/2019-4 (GK), and Funda\c{c}\~{a}o 
de Amparo \`{a} Pesquisa do Estado de S\~{a}o Paulo (FAPESP) process 
Nos.~2019/00763-0 (KT), 64898/2014-5 (KT) and 2013/01907-0 (GK). The work 
is also part of the project Instituto Nacional de Ci\^{e}ncia e Tecnologia 
-- Nuclear Physics and Applications (INCT-FNA), process No.~464898/2014-5 
(KT, GK). It was also supported by the Australian Research Council through 
DP180100497 (AWT).

\end{document}